\documentclass[prb,twocolumn]{revtex4-2}
\usepackage{graphicx} 

\usepackage{amsmath,amsthm,amssymb,mathrsfs}
\usepackage{bm}
\usepackage{color}

\begin{document}

\title{Feedback voltage driven chaos in a three-terminal spin-torque oscillator}

\author{Tomohiro Taniguchi}
\email{tomohiro-taniguchi@aist.go.jp}
\affiliation{
National Institute of Advanced Industrial Science and Technology (AIST), Research Center for Emerging Computing Technologies, Tsukuba, Ibaraki 305-8568, Japan. 
}

\date{\today} 
\begin{abstract}
{
In this work, we report an excitation of chaos and a non-trivial magnetization switching via transient chaos in a three-terminal spin-torque oscillator (STO). 
The driving force of the chaos is a voltage-controlled magnetic anisotropy (VCMA) effect generated by a feedback signal from the STO since the feedback effect is known to be effective in exciting chaos in a dynamical system. 
Solving the Landau-Lifshitz-Gilbert equation numerically and applying temporal and statistical analyses to its solution, the existence of the chaos driven by the feedback VCMA effect is identified. 
Simultaneously, however, transient chaos is also observed, where the magnetization initially shows chaotic behavior but finally switches its direction. 
This transient dynamics from chaos to magnetization switching was unexpected because the sign of the feedback VCMA effect was chosen so that the switching current increases and, as a result, the situation rather favors the condition for sustaining chaos. 
It is implied that this switching happens when narrowing a stable region of the magnetic potential energy by the feedback effect and magnetization precession pointing to a saddle point coincidentally occur simultaneously. 
}
\end{abstract}

\maketitle


\section{Introduction}
\label{sec:Introduction}


Chaos is one kind of nonlinear dynamics which is sensitive to the initial state and thus, is difficult to predict its time evolution. 
These characteristics have attracted attention not only from a fundamental aspect \cite{strogatz01} but also from the viewpoint of practical applications such as random-number generator \cite{stojanovski01}, chaotic communication \cite{cuomo93}, and brain-inspired computing \cite{verstraeten07}. 
While chaos in magnetic systems has been investigated for more than a decade ago by using a relatively large ferromagnet \cite{wigen94,mino01}, recent predictions and observations of chaos in nanostructured ferromagnets suggest a possibility to realize the technology in applications by spintronics devices  \cite{kudo06,li06,yang07,watelot12,devolder19,bondarenko19,montoya19,taniguchi19JMMM,williame19,taniguchi19,yamaguchi19,kamimaki21,taniguchi24PRB}.  
Since chaos can appear only in a high-dimensional phase space \cite{strogatz01,alligood97,ott02}, various methods to increase the number of the dynamical degrees of freedom have been used in the previous works, such as injecting periodic signal \cite{li06,yang07,yamaguchi19} and increasing the number of the ferromagnets \cite{kudo06,taniguchi19JMMM}. 
Among them, adding a feedback signal is an efficient approach to significantly increase the number of the dynamical degree of freedom significantly and to excite chaos \cite{biswas18}. 
A large number of the dynamical degrees of freedom are also preferable for practical applications. 
For example, the number of the linearly independent dynamical degrees of freedom bounds the computational capability of physical reservoir computing \cite{kubota21}. 
Motivated by these reasons, chaos in spin-torque oscillators (STOs) driven by feedback electric current or magnetic field has intensively been investigated both theoretically \cite{williame19,taniguchi19,taniguchi24PRB} and experimentally \cite{kamimaki21,tsunegi23}.  


However, using the feedback electric current or magnetic field has the following undesirable issues. 
A typical STO consists of a magnetic tunnel junction (MTJ), which includes an insulating layer, such as MgO, to detect the magnetization direction by tunnel magnetoresistance effect. 
Therefore, there is a limitation of the amount of electric current to avoid an electro-static breakdown of the MTJ. 
In this respect, the previous works on the feedback electric current focused only on a weak feedback region \cite{khalsa15,tsunegi16}. 
On the other hand, using a feedback magnetic field can avoid this issue because the feedback signal does not directly flow to the MTJ. 
Instead, the feedback signal is sent to a metallic line placed on the STO and generates the feedback magnetic field \cite{kamimaki21,tsunegi23}. 
It is, however, not preferable to use magnetic field for practical applications. 
Considering these points, other ways to make use of feedback effect are greatly anticipated. 
One candidate is the voltage-controlled magnetic anisotropy (VCMA) effect \cite{weisheit07,maruyama09,shiota09}, where an application of voltage to magnetic tunnel junctions modifies magnetic anisotropy energy by a modulation of electron states \cite{duan08,nakamura09,tsujikawa09} and/or induction of magnetic moment \cite{miwa17} near a ferromagnetic metal/insulating barrier interface. 
The VCMA effect does not require additional electric current in principle, and thus, can provide the feedback effect without increasing the total amount of the electric current in the STO. 
In addition, the modulation of the magnetic anisotropy field by the VCMA effect can reach up to on the order of kilo Oersted, which is comparable to or even larger than the Oersted magnetic field used in the previous works \cite{kamimaki21,tsunegi23}.
However, an existence of a feedback signal does not necessarily guarantee an appearance of chaos \cite{biswas18}, and thus, it is still unclear whether the feedback VCMA effect can induce chaos in an STO. 


In this work, we study an excitation of chaos in a three-terminal STO with feedback VCMA effect using numerical simulations of the Landau-Lifshitz-Gilbert (LLG) equation. 
The three-terminal structure enables us to utilize the spin-transfer and VCMA effects simultaneously, where the magnetization oscillation in the STO is excited by spin current generated by the spin Hall effect \cite{dyakonov71,hirsch99,zhang00,kato04} in a bottom electrode while the feedback voltage leading to a bifurcation from simple to chaotic dynamics is applied from the other terminal. 
The existence of chaos caused by the feedback VCMA effect in the three-terminal STO is identified from the temporal solution of the LLG equation and its statistical analyses. 
At the same time, however, transient chaos \cite{lai11} is also observed, where the magnetization initially shows chaotic behavior but finally switches its direction and saturates to a fixed point. 
We notice that this transient dynamics from chaos to magnetization switching is non-trivial because we choose the sign of the feedback VCMA effect so that the switching becomes difficult for sustaining chaos. 
The origin of this transient dynamics is investigated by evaluating the temporal change of the magnetic anisotropy energy due to the feedback VCMA effect. 
The numerical analysis implies that the switching occurs when a stable region of the potential is narrowed by the feedback effect and magnetization precession pointing to a saddle point coincidentally occurs at the same time.



\section{System description}
\label{sec:System description}


\begin{figure}
\centerline{\includegraphics[width=1.0\columnwidth]{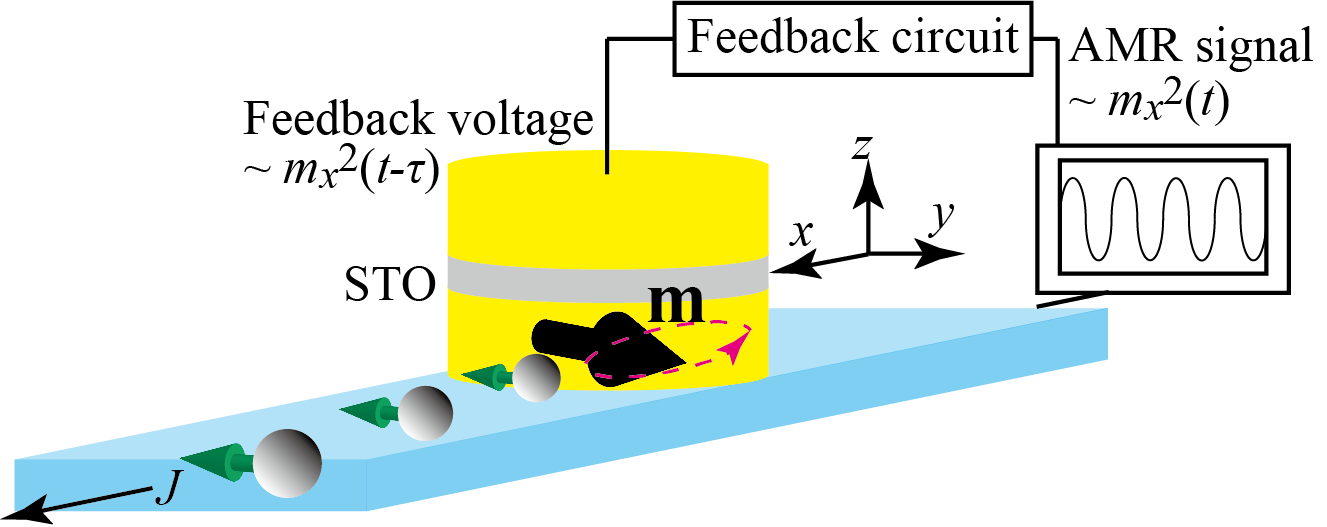}}
\caption{
             Schematic illustration of three-terminal STO with feedback voltage. 
             Electric current density $J$ flowing in the $x$ direction generates spin current polarized in the $y$ direction and is injected in the STO. 
             The spin current excites spin-transfer (spin-orbit) torque on the magnetization $\mathbf{m}$ of a (bottom) free layer in the STO and induces magnetization oscillation around the $y$ axis. 
             The electric current density in the bottom electrode also generates AMR signal ($\propto m_{x}^{2}$), which passes through the feedback circuit with delay time $\tau$ and is converted to feedback voltage applied to the STO from another terminal. 
         \vspace{-3ex}}
\label{fig:fig1}
\end{figure}


Figure \ref{fig:fig1} is a schematic illustration of a three-terminal STO placed on a bottom electrode. 
Electric current density $J$ flowing in the bottom electrode (along the $x$ direction) generates spin current injected into the STO, where the spin is polarized in the $y$ direction. 
The spin current excites spin-transfer \cite{slonczewski96,berger96} (spin-orbit) torque on magnetization in a free layer. 
The magnetization oscillation can be detected by anisotropic magnetoresistance (AMR) effect, as in the case of spin-torque ferromagnetic resonance experiments in three-terminal structures \cite{liu11}.  
Then, this AMR signal is sent to the other terminal through a feedback circuit and is used to determine the feedback voltage, as shown in Fig. \ref{fig:fig1}. 
Therefore, both the spin Hall and VCMA effects coexist in this structure. 
We should mention here that the possibility of utilizing both of them simultaneously was experimentally confirmed in three-terminal magnetoresistive memory \cite{shirotori17}. 
In this work, the magnitude of the voltage for this VCMA effect is determined by the feedback signal.    
The feedback circuit can amplify or attenuate the input signal and yield a delay time $\tau$ \cite{kamimaki21,tsunegi23}, where a typical value of $\tau$ available in experiments is on the order of $1$-$10$ ns. 


Let us mention the advantage and/or necessity using the three-terminal structure, instead of the two-terminal one studied previously \cite{williame19,taniguchi19,kamimaki21,taniguchi24PRB,tsunegi23}. 
The three-terminal structure allows us to simultaneously utilize both the spin-transfer torque and VCMA effects \cite{shirotori17}. 
In the two-terminal device, these effects hardly coexist because their source current and voltage are not independent of each other, and they require contradicting conditions for resistance of the device. 
The spin-transfer torque effect arises from a flow of electrons \cite{slonczewski96,berger96}, while the VCMA effect arises from an accumulation of electrons \cite{nakamura09,tsujikawa09}.  
Thus, the resistances of the spin-transfer torque and VCMA devices are greatly different. 
For example, the resistance of an STO in Ref. \cite{kubota13} is at least one or two orders of magnitude smaller than that of the VCMA devices \cite{shiota17}. 
Accordingly, it is difficult to fabricate devices in which two effects coexist. 
Although small current can flow even in VCMA devices for reading the magnetization direction, it is not sufficient for the current required in STOs applications. 
The three-terminal structure solves this contradiction by separating the paths of the electric current generating the spin-transfer torque and the voltage for the VCMA effect. 
In addition, we assume an in-plane magnetized ferromagnet as a free layer, where an in-plane easy axis is parallel to the direction of the spin-polarization of the spin current. 
This is because this structure enables to excite the auto-oscillation of the magnetization without using an external magnetic field \cite{grollier03}. 
This becomes a key difference when compared with the previous studies on chaos by the VCMA effect, where an external magnetic field is a necessary factor to excite a sustainable dynamics \cite{celada22,taniguchi22,taniguchi24}.


\begin{table}[t]
 \caption{
 Parameters used in the numerical calculations: $M$, saturation magnetization; 
 $H_{\rm K}$, in-plane magnetic anisotropy field;  
 $\gamma$, gyromagnetic ratio;  
 $\alpha$, Gilbert damping constant;  
 $d$, thickness of the free layer; 
 $\vartheta$, spin Hall angle; 
 $J$, current density; 
 $\tau$, delay time. 
 }
 \label{table:table1}
 \centering
  \begin{tabular}{c c}
   \hline \hline
   Quantity & Value\\
   \hline 
   $M$ & 1000 emu/cm$^{3}$ \\
   $H_{\rm K}$ & 200 Oe \\
   $\gamma$ & $1.764\times 10^{7}$ rad/(Oe s) \\ 
   $\alpha$ & $0.010$ \\
   $d$ & $2$ nm \\
   $\vartheta$ & 0.30 \\
   $J$ & $15.0$ MA/cm${}^{2}$ \\
   $\tau$ & $20$ ns \\
         \hline
  \end{tabular}
\end{table}


The magnetization dynamics in the proposed STO is described by the LLG equation given by 
\begin{equation}
  \frac{d\mathbf{m}}{dt}
  =
  -\gamma
  \mathbf{m}
  \times
  \mathbf{H}
  -
  \frac{\gamma\hbar\vartheta J}{2eMd}
  \mathbf{m}
  \times
  \left(
    \mathbf{e}_{y}
    \times
    \mathbf{m}
  \right)
  +
  \alpha
  \mathbf{m}
  \times
  \frac{d\mathbf{m}}{dt},
  \label{eq:LLG}
\end{equation}
where $\mathbf{m}=(m_{x},m_{y},m_{z})$ is the unit vector pointing in the direction of the magnetization in the free layer, while $\mathbf{e}_{k}$ ($k=x,y,z$) is the unit vector in the $k$ direction. 
The gyromagnetic ratio, saturation magnetization, thickness along the $z$ direcction, and the Gilbert damping constant of the free layer are denoted as $\gamma$, $M$, $d$, and $\alpha$, respectively. 
The spin Hall angle is $\vartheta$. 
We assume that the free layer in the STO is in-plane magnetized and its magnetic field $\mathbf{H}$ is given by 
\begin{equation}
  \mathbf{H}
  =
  H_{\rm K}
  m_{y}
  \mathbf{e}_{y}
  -
  4\pi M 
  \left\{
    1
    +
    \nu 
    [m_{x}(t-\tau)]^{2}
  \right\}
  m_{z}
  \mathbf{e}_{z},
  \label{eq:field}
\end{equation}
where $H_{\rm K}$ is the in-plane magnetic anisotropy field along the $y$ direction while $4\pi M$ is the demagnetization field along the $z$ direction. 
The term $\nu [m_{x}(t-\tau)]^{2}$ represents the modulation of the perpendicular magnetic anisotropy field by the VCMA effect, where the magnitude of the modulation is determined by the AMR feedback signal proportional to $m_{x}^{2}$. 
The dimensionless coefficient $\nu$ is defined as $H_{{\rm K},V}/(4\pi M)$, where $H_{{\rm K},V}=2\eta V/(Mdd_{\rm I})$ represents the magnitude of the VCMA effect in terms of magnetic field with the VCMA efficiency $\eta$, voltage $V$, and the thickness $d_{\rm I}$ of the insulating layer in the STO. 
The VCMA efficiency recently obtained in the experiment reaches about $350$ fJ/(V m) \cite{nozaki20}, which makes $H_{{\rm K},V}$ on the order of kilo Oersted for typical device structures. 
The magnitude of the voltage, $|V|$, which possibly includes an amplification or attenuation by the feedback circuit, depends on its pulse width to avoid electrostatic breakdown of the STO.
It can be about $3.0$ V for an ultra-short pulse but is limited to $1.0$ V or less at maximum for a relatively long or continuous application of the voltage. 
The saturation magnetization $M$ in typical STOs and/or VCMA devices are about $1000$ emu/cm${}^{3}$. 
Regarding these values, the magnitude of $\nu$ can be on the order of $0.1$ (or equivalently, $H_{{\rm K},V}$ can be on the order of kilo Oersted). 
The sign of $\nu$ can be either positive or negative, depending on the sign of the voltage. 
In the following, we focus on the positive $\nu$. 
The reason will be explained in Sec. \ref{sec:Analysis of transient chaos}. 
The values of the parameters used in the following calculations, except $\nu$, are summarized in Table \ref{table:table1}. 
We solved Eq. (\ref{eq:LLG}) with an initial condition $\mathbf{m}(0)=(\sin90^{\circ}\cos89^{\circ},\sin90^{\circ}\sin89^{\circ},\cos90^{\circ})$ from $t=0$ to $t=t_{\rm max}$, where $t_{\rm max}=1.0$ $\mu$s throughout this work. 
The choice of the initial condition does not affect whether chaos occurs or not, although the time evolution of the magnetization depends on it; see also Sec. \ref{sec:Discussion on numerical results}. 
In addition, as a general consequence for chaos and transient chaos \cite{lai11}, the results shown below, such as whether the systems shows chaos or transient chaos, depend on the choice of $t_{\rm max}$, which will be discussed later.  
The fourth-order Runge-Kutta method extended to feedback system \cite{hairer93,taniguchi19} was applied for solving Eq. (\ref{eq:LLG}), where the time increment of the LLG equation is defined from the delay time $\tau$ and an integer $N$ as $\Delta t=\tau/N$ ($N=2\times 10^{4}$, or equivalently, $\Delta t=1$ ps, in this work). 
The feedback term is zero until $t<\tau$ because before applying electric current there is no output signal sent to the feedback circuit.  
The value of the time increment, $\Delta t$, is chosen so that it should be much shorter than characteristic time scales of the present system such as the precession period and switching time of the magnetization and the inverse of the Lyapunov exponent, which are shown in the next section. 


For the later discussion, we note that the number of the dynamical degree of the freedom in this numerical scheme is $2(N+1)$ \cite{hairer93,taniguchi19} due to the following reason. 
Recall that the number of independent dynamical variables in the conventional macrospin system is two because, while $\mathbf{m}$ is a vector in a three-dimensional space, there is a constraint $|\mathbf{m}|=1$ in the LLG equation. 
Once its initial condition $\mathbf{m}(0)$ is given, the time evolution of $\mathbf{m}(t)$ is uniquely determined by the LLG equation. 
On the other hand, in the feedback system, an infinite number of the initial conditions is necessary because the time evolution from the initial time $t=0$ to $t=\tau$ depends on the values of the feedback terms in the LLG equation; thus, in principle (or mathematically), the dimension of the feedback system is infinite \cite{biswas18}. 
When solving the LLG equation numerically by the numerical method mentioned above, the discretization of time makes the number of independent initial conditions finite, $N+1$. 
Summarizing them, the total dimension of the present system is $2(N+1)$. 
This fact makes not only the calculation costs heavy but also the evaluation of some characteristics, such as Lyapunov exponent, complex, as will be mentioned in Sec. \ref{sec:Statistical analysis of Lyapunov exponent} below. 



\section{Numerical results}
\label{sec:Numerical results}

In this section, we summarize the results of the numerical simulations. 


\begin{figure*}
\centerline{\includegraphics[width=2.0\columnwidth]{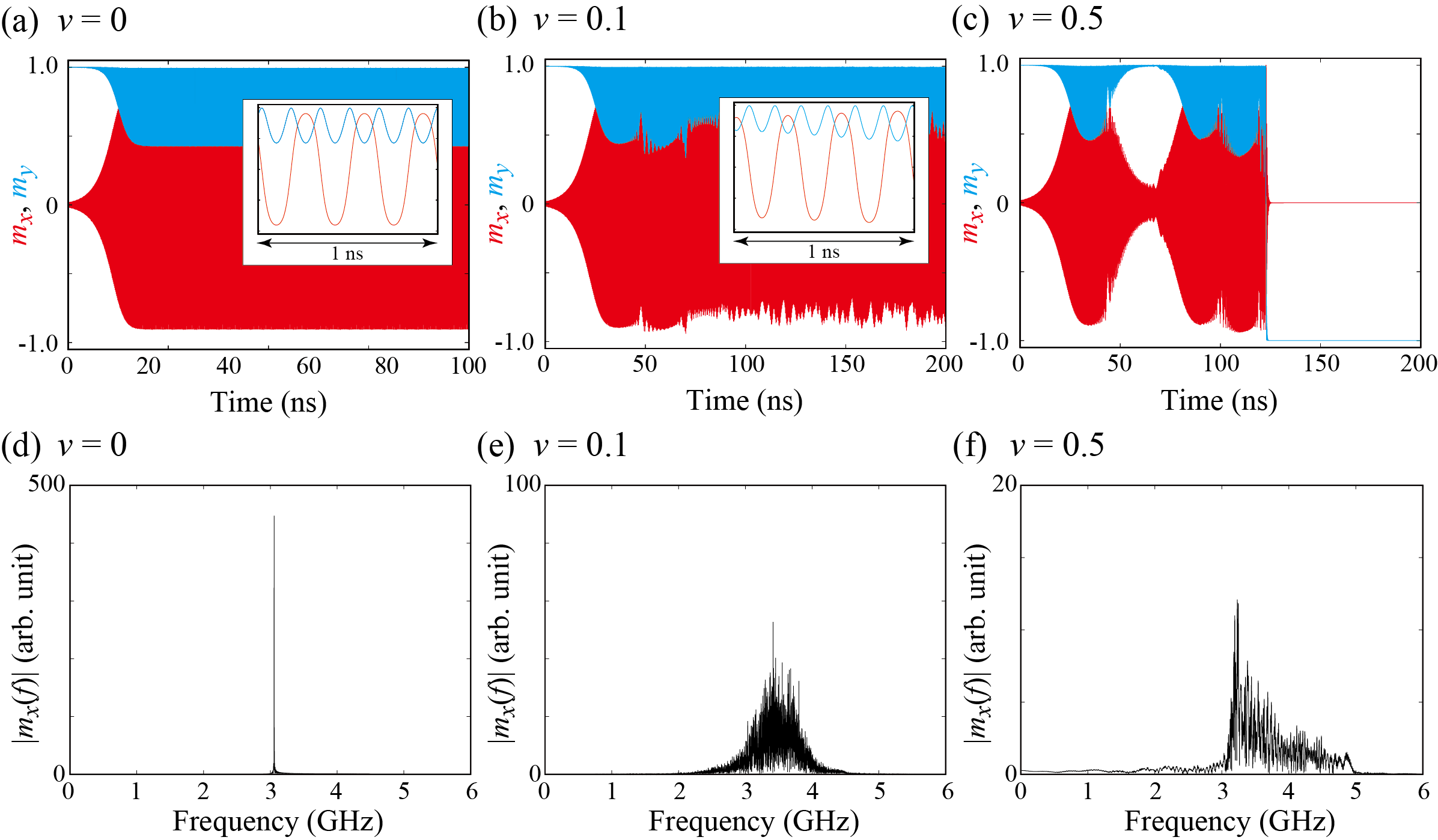}}
\caption{
            Temporal dynamics of $m_{x}$ and $m_{y}$ for $\nu$ of (a) $0$, (b) $0.1$, and (c) $0.5$. 
            The insets in (a) and (b) show their dynamics in a time range of $1$ ns. 
            Fourier spectra, $|m_{x}(f)|$, of $m_{x}(t)$ for those $\nu$'s are shown in (d)-(f). 
         \vspace{-3ex}}
\label{fig:fig2}
\end{figure*}


\subsection{Temporal dynamics}
\label{sec:Temporal dynamics}

\begin{figure}
\centerline{\includegraphics[width=1.0\columnwidth]{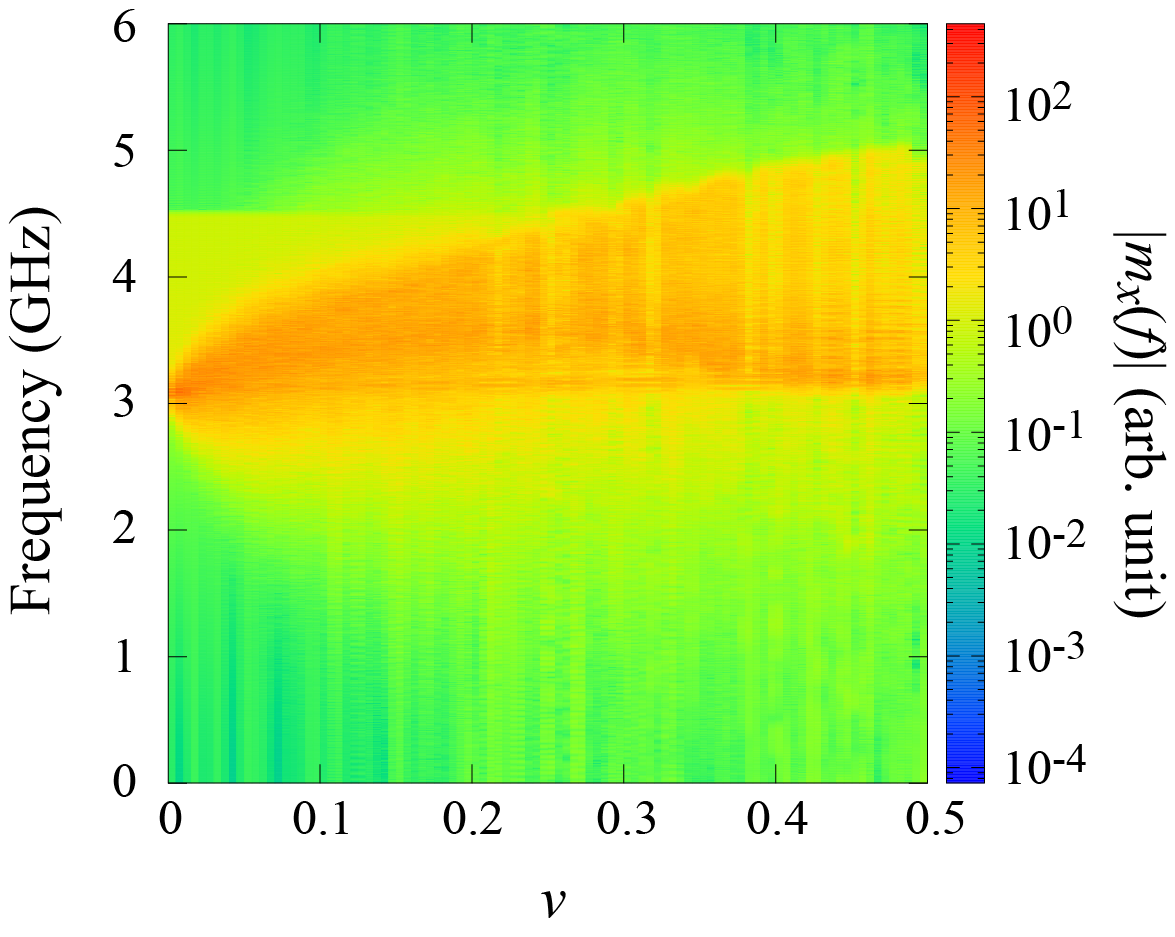}}
\caption{
            Power spectrum density, $|m_{x}(f)|$, for various $\nu$'s, where $m_{x}(f)$ is the Fourier transformation of $m_{x}(t)$. 
         \vspace{-3ex}}
\label{fig:fig3}
\end{figure}

Figures \ref{fig:fig2}(a)-\ref{fig:fig2}(c) summarize typical dynamics of $m_{x}$ (red) and $m_{y}$ (blue) near $t=0$, where $\nu$ is (a) $0$, (b) $0.1$, and (c) $0.5$, while the Fourier spectra, $|m_{x}(f)|$, of $m_{x}(t)$ for these $\nu$ are shown in Figs. \ref{fig:fig2}(d)-\ref{fig:fig2}(f). 
The Fourier spectra for various $\nu$ are summarized in Fig. \ref{fig:fig3}. 
In the absence of the feedback effect ($\nu=0$), an auto-oscillation with a single main frequency is excited, as in the case of the conventional three-terminal STOs \cite{liu12}; see Figs. \ref{fig:fig2}(a) and \ref{fig:fig2}(d), as well as the inset in Fig. \ref{fig:fig2}(a). 
In the presence of the feedback effect, however, the dynamics is largely modulated, although the dynamics during a short time scale looks similar to a simple auto-oscillation; see Fig. \ref{fig:fig2}(b) and compare the insets in Fig. \ref{fig:fig2}(a) and \ref{fig:fig2}(b).  
The complexity of the dynamics is also found from the Fourier spectrum in Fig. \ref{fig:fig2}(e), where the spectra shows broadened structure. 
The magnetization switching is also observed for different values of $\nu$, as shown in Fig. \ref{fig:fig2}(c). 
Note that the oscillation amplitude repeats the cycles of amplification and reduction before the switching, which is also reflected in the broadened structure of the Fourier spectrum in Fig. \ref{fig:fig2}(f). 
This switching behavior is quite different from the conventional spin-transfer (spin-orbit) torque switching \cite{sun00,grollier03}, where the work done by the spin-transfer torque monotonically supplies energy into the ferromagnet and thus, the oscillation amplitude increases monotonically. 
The complex behavior before the magnetization switching found in Fig. \ref{fig:fig2}(c) suggests that the dynamics is classified as transient chaos \cite{lai11}, where the nonlinear dynamical system shows chaotic behavior before saturating to a fixed point. 


Summarizing the results, it is shown that the feedback VCMA effect causes the modulation of the oscillation amplitude in the STO, and the dynamics becomes complex. 
The complexity of the dynamics results in the broadened structure of the Fourier spectra, as summarized in Fig. \ref{fig:fig3}. 
These results are similar to the characteristics found in chaotic spintronics devices with feedback current or magnetic field \cite{taniguchi19,kamimaki21,taniguchi24PRB}. 
We should, however, recall that an appearance of complex dynamics is not a direct evidence of the existence of chaos. 
Although the complexity of dynamics is a typical characteristic of chaos, not all complex dynamics is classified as chaos. 
One way to identify chaos is to evaluate the Lyapunov exponent \cite{strogatz01}. 


\subsection{Statistical analysis of Lyapunov exponent}
\label{sec:Statistical analysis of Lyapunov exponent}

The Lyapunov exponent is an instantaneous expansion rate of a distance between two solutions of the LLG equation having infinitesimally different initial conditions. 
The expansion rate is usually time and/or position dependent, and thus, a statistical average of temporal Lyapunov exponent is often evaluated \cite{shimada79,farmer82,sano85,rosenstein93,kanno14}. 
The Lyapunov exponent in spintronics system has been studied previously \cite{li06,yang07,yamaguchi19,celada22,taniguchi22,taniguchi24}. 
However, a different method is necessary for evaluating the Lyapunov exponent in the feedback system because not only the expansion rate of the magnetization at the present time but also that in the past time should be evaluated. 
Therefore, let us first explain the evaluation method of the Lyapunov exponent in the present system. 

Let us denote the solution of the LLG equation with an initial condition $\mathbf{m}(t)$. 
Note that we need to store the value of $N+1$ $\mathbf{m}$ [$\mathbf{m}(t-\tau)$, $\mathbf{m}(t-\tau+\Delta t)$, $\mathbf{m}(t-\tau+2\Delta t)$, $\cdots$, $\mathbf{m}(t-\Delta t)$, $\mathbf{m}(t)$] for performing the numerical simulation, where $\mathbf{m}(t-\tau), \cdots, \mathbf{m}(t-\Delta t)$ are the past values of $\mathbf{m}(t)$ and are used as feedback signal. 
Assuming that we start to evaluate the Lyapunov exponent from a certain time $t=t_{\rm init}$, let us denote the set of these vectors as 
\begin{equation}
  \mathbf{Y}(t_{\rm init})
  =
  \begin{pmatrix} 
    \mathbf{m}(t_{\rm init}-\tau) \\
    \mathbf{m}(t_{\rm init}-\tau+\Delta t) \\ 
    \vdots \\
    \mathbf{m}(t_{\rm init})
  \end{pmatrix}. 
\end{equation}
Their values are determined by the LLG equation with the initial condition mentioned in Sec. \ref{sec:System description}. 
These vectors are the dynamical variables in the feedback system \cite{biswas18}. 
Since each vector $\mathbf{m}$ is a three-dimensional vector with a constraint $|\mathbf{m}|=1$, the number of the dynamical degree of freedom is $2(N+1)$, as mentioned earlier.    
We note that the ordering of these $N+1$ $\mathbf{m}$ does not affect the evaluation of the distance in the phase space, as well as the Lyapunov exponent, by the method mentioned below and thus, is arbitrary.
To evaluate the Lyapunov exponent, for each $\mathbf{m}[t_{\rm init}-(N-k) \Delta t]$ ($k=0,1,\cdots,N$), we introduce the zenith and azimuth angles in a spherical coordinate, $\theta_{k}=\cos^{-1}m_{z}[t_{\rm init}-(N-k)\Delta t]$ and $\varphi_{k}=\tan^{-1}\{m_{y}[t_{\rm init}-(N-k)\Delta t]/m_{x}[t_{\rm init}-(N-k)\Delta t]\}$. 
Then, we introduce $N+1$ $\mathbf{m}_{1}$, whose zenith and azimuth angles, $\theta_{1,k}=\cos^{-1}m_{1,k,z}[t_{\rm init}-(N-k) \Delta t]$ and $\varphi_{1,k}=\tan^{-1}\{m_{1,k,y}[t_{\rm init}-(N-k)\Delta t]/m_{1,k,x}[t_{\rm init}-(N-k)\Delta t]\}$, are defined by shifting $\theta_{k}$ and $\varphi_{k}$ with infinitesimal differences ($|\epsilon_{1,k,\theta}|,|\epsilon_{1,k,\varphi}|\ll 1$) as $\theta_{1,k}=\theta_{k}+\epsilon_{1,k,\theta}$ and $\varphi_{1,k}=\varphi_{k}+\epsilon_{1,k,\varphi}$. 
Here, the subscript ``1'' is applied to the variables necessary to evaluate the first temporal Lyapunov exponent defined below.
The distance between $\mathbf{m}[t_{\rm init}-(N-k)\Delta t]$ and $\mathbf{m}_{1}[t_{\rm init}-(N-k)\Delta t]$ at time $t_{\rm init}$ is given by $\epsilon_{1,k}=\sqrt{\epsilon_{1,k,\theta}^{2}+\epsilon_{1,k,\varphi}^{2}}$, and their sum is $\epsilon=\sum_{k=0}^{N}\epsilon_{1,k}$. 
Also, we introduce 
\begin{equation}
  \mathbf{Y}_{1}(t_{\rm init})
  =
  \begin{pmatrix}
    \mathbf{m}_{1}(t_{\rm init}-\tau) \\ 
    \mathbf{m}_{1}(t_{\rm init}-\tau+\Delta t) \\ 
    \vdots \\
    \mathbf{m}_{1}(t_{\rm init})
  \end{pmatrix}.
\end{equation} 
Therefore, $\epsilon$ can be regarded as a distance between $\mathbf{Y}(t_{\rm init})$ and $\mathbf{Y}_{1}(t_{\rm init})$. 
Then, we evaluate the time evolution of $\mathbf{m}(t_{\rm init})$ and $\mathbf{m}_{1}(t_{\rm init})$ from time $t_{\rm init}$ to $t_{\rm init}+\Delta t$ by solving the LLG equation and obtain $\mathbf{m}(t_{\rm init}+\Delta t)$ and $\mathbf{m}_{1}(t_{\rm init}+\Delta t)$. 
Recall that the values of $\mathbf{m}(t_{\rm init}-\tau)$ and $\mathbf{m}_{1}(t_{\rm init}-\tau)$ in $\mathbf{Y}(t_{\rm init})$ and $\mathbf{Y}_{1}(t_{\rm init})$, respectively, are used as the feedback term to evaluate the time evolution and become no longer necessary to be stored. 
Thus, we update $\mathbf{Y}$ and $\mathbf{Y}_{1}$ by replacing $\mathbf{m}(t_{\rm init}-\tau)$ and $\mathbf{m}_{1}(t_{\rm init}-\tau)$ in them respectively with $\mathbf{m}(t_{\rm init}+\Delta t)$ and $\mathbf{m}_{1}(t_{\rm init}+\Delta t)$ as 
\begin{equation}
  \mathbf{Y}(t_{\rm init}+\Delta t)
  =
  \begin{pmatrix}
    \mathbf{m}(t_{\rm init}+\Delta t) \\ 
    \mathbf{m}(t_{\rm init}-\tau+\Delta t) \\ 
    \vdots \\ 
    \mathbf{m}(t_{\rm init})
  \end{pmatrix},
\end{equation}
\begin{equation}
  \mathbf{Y}_{1}(t_{\rm init}+\Delta t)
  =
  \begin{pmatrix}
    \mathbf{m}_{1}(t_{\rm init}+\Delta t) \\
    \mathbf{m}_{1}(t_{\rm init}-\tau+\Delta t) \\ 
    \vdots \\
    \mathbf{m}_{1}(t_{\rm init})
  \end{pmatrix}.
\end{equation} 
Note that the distance $\epsilon_{1}$ between $\mathbf{Y}(t_{\rm init}+\Delta t)$ and $\mathbf{Y}_{1}(t_{\rm init}+\Delta t)$ is different from that ($\epsilon$) between $\mathbf{Y}(t_{\rm init})$ and $\mathbf{Y}_{1}(t_{\rm init})$. 
They are related as $\epsilon_{1}=\epsilon-\epsilon_{1,0}+\epsilon_{1,0}^{\prime}$, where $\epsilon_{1,0}^{\prime}$ is the distance between $\mathbf{m}(t_{\rm init}+\Delta t)$ and $\mathbf{m}_{1}(t_{\rm init}+\Delta t)$, in terms of $\theta_{0}^{\prime}=\cos^{-1}m_{z}(t_{\rm init}+\Delta t)$, $\varphi_{0}^{\prime}=\tan^{-1}[m_{y}(t_{\rm init}+\Delta t)/m_{x}(t_{\rm init}+\Delta t)]$, $\theta_{1,0}^{\prime}=\cos^{-1}m_{1,z}(t_{\rm init}+\Delta t)$, and $\varphi_{1,0}^{\prime}=\tan^{-1}[m_{1,y}(t_{\rm init}+\Delta t)/m_{1,x}(t_{\rm init}+\Delta t)]$. 
The expansion rate of the solution of the LLG equation during $[t_{\rm init},t_{\rm init}+\Delta t]$ is given by $\epsilon_{1}/\epsilon$. 
Therefore, a temporal Lyapunov exponent at time $t_{\rm init}+\Delta t$ is defined as 
\begin{equation}
  \lambda_{1}
  =
  \frac{1}{\Delta t}
  \ln
  \frac{\epsilon_{1}}{\epsilon}. 
\end{equation}
We repeat this procedure and evaluate the temporal Lyapunov exponent every time (in this work, we choose $t_{\rm init}=\tau$), i.e., we start to evaluate the temporal Lyapunov exponent from $t=\tau$). 
In general, at time $t_{n}=t_{\rm init}+(n-1)\Delta t$ ($n=1,2,\cdots$), we introduce $\mathbf{Y}_{n}(t_{n})$, which consists of $N+1$ $\mathbf{m}$, $\mathbf{m}_{n}(t_{n}-\tau)$, $\mathbf{m}_{n}(t_{n}-\tau+\Delta t)$, $\cdots$, $\mathbf{m}_{n}(t_{n})$, and differs from $\mathbf{Y}(t_{n})$ with a distance $\epsilon$, i.e., $|\mathbf{Y}(t_{n})-\mathbf{Y}_{n}(t_{n})|=\epsilon$. 
Here, $\mathbf{m}_{n}(t_{n}-\tau+k\Delta t)$ is defined from $\mathbf{m}(t_{n}-\tau+k\Delta t)$ so that the direction from $\mathbf{m}(t_{n}-\tau+k\Delta t)$ to $\mathbf{m}_{n}(t_{n}-\tau+k\Delta t)$ is the most expanded direction; see Refs. \cite{taniguchi19,taniguchi24,shimada79}.
Then, the time evolution of $\mathbf{m}(t_{n})$ and $\mathbf{m}_{n}(t_{n})$ to $\mathbf{m}(t_{n}+\Delta t)$ and $\mathbf{m}_{n}(t_{n}+\Delta t)$ are evaluated, and $\mathbf{Y}(t_{n})$ and $\mathbf{Y}_{n}(t_{n})$ are updated to $\mathbf{Y}(t_{n}+\Delta t)$ and $\mathbf{Y}_{n}(t_{n}+\Delta t)$, respectively, by replacing $\mathbf{m}(t_{n}-\tau)$ and $\mathbf{m}_{n}(t_{n}-\tau)$ in them with $\mathbf{m}(t_{n}+\Delta t)$ and $\mathbf{m}_{n}(t_{n}+\Delta t)$. 
From their distance $\epsilon_{n}=|\mathbf{Y}_{n}(t_{n}+\Delta t)-\mathbf{Y}(t_{n}+\Delta t)|$, the $n$th temporal Lyapunov exponent is defined as $\lambda_{n}=(1/\Delta t)\ln (\epsilon_{n}/\epsilon)$. 
After that, we introduce $\mathbf{Y}_{n+1}(t_{n}+\Delta t)$ from $\mathbf{Y}(t_{n}+\Delta t)$, where the elements in $\mathbf{Y}_{n+1}(t_{n}+\Delta t)$ are determined by moving those in $\mathbf{Y}(t_{n}+\Delta t)$ to the most expanded direction with the condition $|\mathbf{Y}(t_{n}+\Delta t)-\mathbf{Y}_{n+1}(t_{n}+\Delta t)|=\epsilon$, as we define $\mathbf{Y}_{n}(t_{n})$ mentioned above.  
Note that the most expanded direction is determined by $\mathbf{Y}(t_{n}+\Delta t)$ and $\mathbf{Y}_{n}(t_{n}+\Delta t)$; see Refs. \cite{taniguchi19,taniguchi24,shimada79} in detail. 
Finally, the Lyapunov exponent $\varLambda$ is defined as a statistical average of the temporal Lyapunov exponents as $\varLambda=(1/\mathscr{N})\sum_{n=1}^{\mathscr{N}}\lambda_{n}$ \cite{taniguchi19}, where $\mathscr{N}$ is the number of the averaging and $\lambda_{n}$ is the $n$th temporal Lyapunov exponent 
(since we prepare $\mathbf{m}_{1}$ at $t=\tau$, as mentioned above, and the time increment $\Delta t=1$ ps while $t_{\rm max}=1$ $\mu$s, $\mathscr{N}=979999$ in this work).  

\begin{figure}
\centerline{\includegraphics[width=1.0\columnwidth]{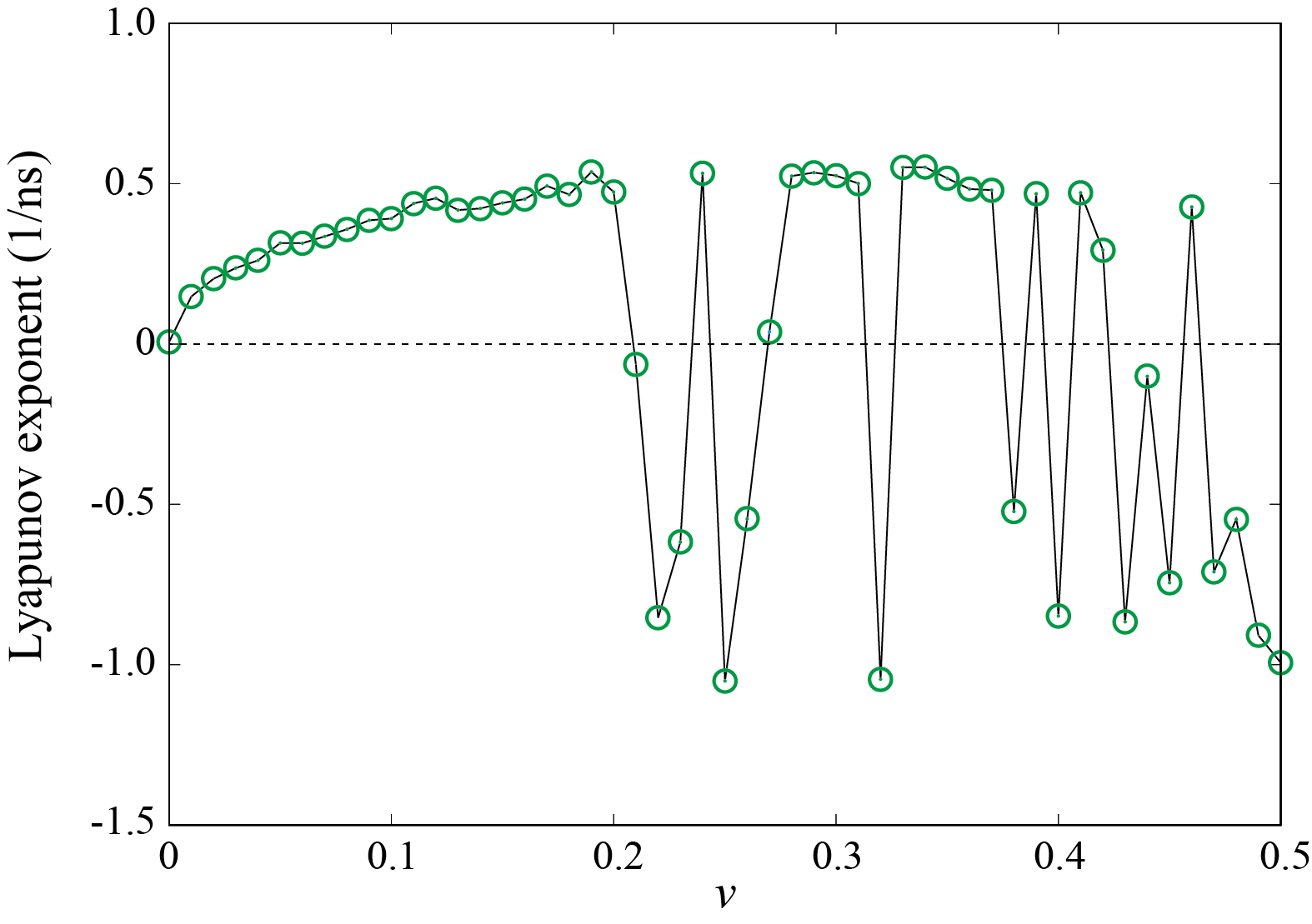}}
\caption{
            Lyapunov exponent for various $\nu$. 
            A horizontal dotted line is added as a guide for eyes to clarify a boundary of zero exponent. 
         \vspace{-3ex}}
\label{fig:fig4}
\end{figure}

The sign of the Lyapunov exponent indicates the dynamical state of the system, i.e., negative exponent corresponds to dynamics saturating to a fixed point, zero exponent corresponds to a periodic dynamics, and positive exponent corresponds to chaos. 
Figure \ref{fig:fig4} summarizes the Lyapunov exponent for various $\nu$'s. 
The value is zero at $\nu=0$, i.e., in the absence of the feedback effect, because the magnetization dynamics in this case is a periodic oscillation (auto-oscillation), as shown in Fig. \ref{fig:fig2}(a). 
For finite feedback effect ($\nu \neq 0$), such as the dynamics in Fig. \ref{fig:fig2}(b), the Lyapunov exponent becomes positive, which can be an evidence of the existence of chaos in the present system. 
The Lyapunov exponent can also be negative for some $\nu$'s when the magnetization switches its direction, as shown in Fig. \ref{fig:fig2}(c). 
Summarizing them, the sign of the Lyapunov exponent reasonably reflects the dynamical behavior of the magnetization. 



\subsection{Discussion on numerical results}
\label{sec:Discussion on numerical results}

Note that analyses on both temporal dynamics in Fig. \ref{fig:fig2} and statistical property (Lyapunov exponent) in Fig. \ref{fig:fig4} are necessary to identify dynamical state of the magnetization. 
The temporal analysis clarifies the existence of complex dynamics, at least as temporary, which may be failed to notice from just evaluating the Lyapunov exponent. 
For example, for the dynamics shown in Fig. \ref{fig:fig2}(c), chaotic behavior that appeared initially might be overlooked when only the Lyapunov exponent is evaluated. 
On the other hand, the evaluation of the Lyapunov exponent can be used as a figure of merit to distinguish between chaos and complex but non-chaotic dynamics, which is hard to determine from just observing temporal dynamics. 
While the temporal dynamics in STOs has been frequently measured and used in practical applications, one may be interested in how to estimate the Lyapunov exponent experimentally. 
In experiments, it is difficult to control the initial state of the magnetization, and thus, the evaluation method of the Lyapunov exponent used in the numerical simulation is hardly applicable. 
There are, however, various methods to estimate the Lyapunov exponent from experimental data by reproducing the dynamical trajectory in an embedded space \cite{alligood97} or identify chaos from different quantities, such as noise limit \cite{barahona96,poon01,devolder19,kamimaki21}. 
Therefore, it will be possible to examine the validity of the present results by experiments in the future. 


Recall that the identification of chaos sometimes depends on the measurement time $t_{\rm max}$, especially when transient chaos possibly occurs \cite{lai11}. 
For example, we did not observe magnetization switching until $t_{\rm max}=1$ $\mu$s for $\nu=0.1$ shown in Fig. \ref{fig:fig2}(b); therefore, the corresponding Lyapunov exponent for this $\nu$ is positive, as shown in Fig. \ref{fig:fig4}. 
If we use, however, $t_{\rm max}$ longer than $1$ $\mu$s, a switching may occur; in this case, the Lyapunov exponent will be negative and the dynamics will be classified as transient chaos. 
On the other hand, if we, for example, use $t_{\rm max}=100$ ns, the dynamics for $\nu=0.5$ shown in Fig. \ref{fig:fig2}(c) may be classified as chaos and the Lyapunov exponent becomes positive because at $t=100$ ns, there is no switching. 
As shown in these examples, the identification of chaos depends on the measurement time. 
This is not a particular property for spintronics devices; rather, this is a general consequence in chaotic system \cite{lai11}. 
It is generally difficult to estimate the transient time, if transient chaos exists, because it depends not only on the values of parameters but also on the initial conditions, \textit{etc} \cite{lai11}. 
The time scale during which chaos can be kept determines the manipulation time of chaos applications, and thus, is of great interest. 
We keep this issue as a future work. 


\section{Analysis of transient chaos}
\label{sec:Analysis of transient chaos}

The numerical simulations revealed that the feedback VCMA effect results not only in chaos but also in transient chaos \cite{lai11}. 
In the transient chaos, the magnetization initially shows chaotic behavior but finally switches its direction and saturates to a fixed point. 
We note that the appearance of this transient dynamics is unexpected and non-trivial. 
In this section, we explain the reason why this switching is non-trivial and investigate its origin. 



\begin{figure*}
\centerline{\includegraphics[width=2.0\columnwidth]{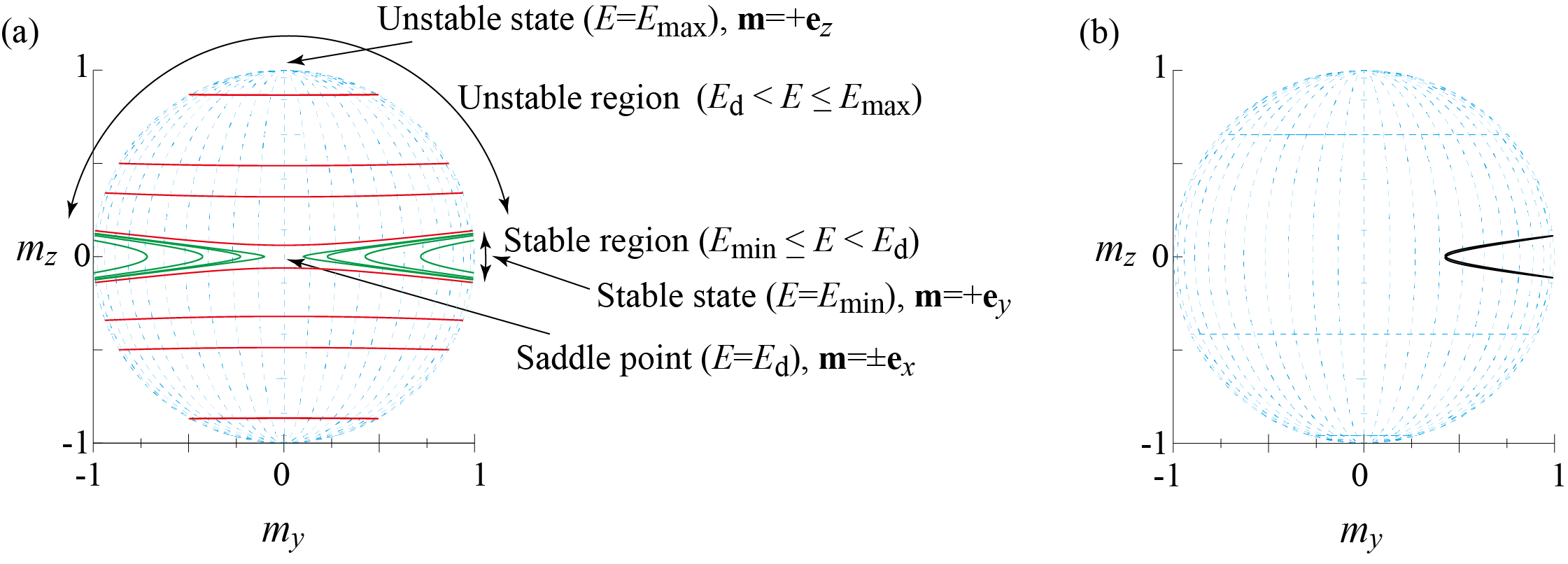}}
\caption{
            (a) Examples of constant energy lines of $E$. 
                The lines belong to either stable or unstable region, depending on its value. 
                We call the points $\mathbf{m}=\pm\mathbf{e}_{y}(\pm\mathbf{e}_{z})$ stable (unstable) states because these points correspond to the energy minima (maxima). 
                The points $\mathbf{m}=\pm\mathbf{e}_{x}$ are the saddle points. 
                While only one stable and unstable regions, around $\mathbf{m}=+\mathbf{e}_{y}$ and $\mathbf{m}=+\mathbf{e}_{z}$, are shown in the figure for simplicity, there are other stable and unstable regions around $\mathbf{m}=-\mathbf{e}_{y}$ and $\mathbf{m}=-\mathbf{e}_{z}$.
            (b) Dynamical trajectory in an auto-oscillation state shown in Fig. \ref{fig:fig2}(a). 
         \vspace{-3ex}}
\label{fig:fig5}
\end{figure*}


\subsection{Why is the switching non-trivial?}
\label{sec:Why is the switching non-trivial?}

For a while, let us neglect the feedback effect in the demagnetization field, for simplicity. 
The previous works clarified that there are two characteristic current densities for the auto-oscillation of the magnetization in an in-plane magnetized ferromagnet. 
One is a critical current density $J_{\rm c}$ given by \cite{sun00,grollier03,bazaliy01,bazaliy03,bazaliy04}, 
\begin{equation}
  J_{\rm c}
  =
  \frac{2\alpha eMd}{\hbar\vartheta}
  \left(
    H_{\rm K}
    +
    2\pi M
  \right), 
\end{equation}
while the other is a threshold current density $J^{*}$ given by \cite{hillebrands06,bazaliy07,taniguchi13,taniguchi14JAP}, 
\begin{equation}
  J^{*}
  =
  \frac{4\alpha eMd}{\pi\hbar\vartheta}
  \sqrt{
    4\pi M 
    \left(
      H_{\rm K}
      +
      4\pi M 
    \right)
  }.
\end{equation}
These current densities relate to the stability of the magnetization having a magnetic potential energy density $E$, which relates to the magnetic field $\mathbf{H}$ via $\mathbf{H}=-\partial E/\partial (M \mathbf{m})$ and is given by 
\begin{equation}
  E
  =
  -\frac{MH_{\rm K}}{2}
  m_{y}^{2}
  +
  2\pi M^{2}
  m_{z}^{2}.
  \label{eq:energy}
\end{equation}
The energy density $E$ has two minima ($E_{\rm min}=-MH_{\rm K}/2$) at $\mathbf{m}=\pm \mathbf{e}_{y}$, two saddle points ($E_{\rm d}=0$) at $\mathbf{m}=\pm\mathbf{e}_{x}$, and two maxima ($E_{\rm max}=2\pi M^{2}$) at $\mathbf{m}=\pm \mathbf{e}_{z}$. 
Recall that a constant energy curve of $E$ is identical to a solution of Landau-Lifshitz (LL) equation, $d\mathbf{m}/dt=-\gamma\mathbf{m}\times\mathbf{H}$, which is the LLG equation without spin-transfer and damping torques. 
In Fig. \ref{fig:fig5}(a), we show examples of constant energy lines. 
Since the LLG equation conserves the magnitude of the magnetization ($|\mathbf{m}|=1$), the constant energy lines can be represented as lines on a unit sphere. 
Let us call the region satisfying $E_{\rm min}\le E<E_{\rm d}$ ($E_{\rm d}< E \le E_{\rm max}$) stable (unstable) region. 
In particular, the points $\mathbf{m}=\pm\mathbf{e}_{y}(\pm\mathbf{e}_{z}$) are called stable (unstable) states. 
Note that the constant energy lines in the stable region are densely distributed due to the large demagnetization field.
It should be noted that the trajectory of the conventional auto-oscillation state is well approximated by a constant energy line. 
This is because the existence of a sustainable oscillation means that the spin-transfer (spin-orbit) and damping torques balance each other and thus, $E$ is approximately kept to be constant; see also Appendix \ref{sec:AppendixA}. 
For example, in Fig. \ref{fig:fig5}(b), we show the steady-state dynamics in Fig. \ref{fig:fig2}(a) as a trajectory, which is similar to a constant energy line in Fig. \ref{fig:fig5}(a). 
Note also that this oscillation trajectory stays inside the stable region in Fig. \ref{fig:fig5}(a), which means that the energy in an auto-oscillation state is in the range of $E_{\rm min}<E<E_{\rm d}$. 


The critical current density $J_{\rm c}$ is derived from the LLG equation linearized near the stable state and determines the stability around it; see also Appendix \ref{sec:AppendixA}. 
On the other hand, the threshold current density $J^{*}$ is derived from the stability condition of the LLG equation averaged over a constant energy line including the saddle points \cite{taniguchi13}; see also Appendix \ref{sec:AppendixA}.  
The auto-oscillation of the magnetization occurs when the current density $J$ is in the range of $J_{\rm c}<J<J^{*}$ (see also Appendix \ref{sec:AppendixA}), whereas the magnetization switching occurs when $J>J^{*}$. 
In the present system, the values of $J_{\rm c}$ and $J^{*}$ in the absence of the feedback effect are $13.1$ MA/cm${}^{2}$ and $16.3$ MA/cm${}^{2}$, respectively. 
Recall that the current density $J$ was $15.0$ MA/cm${}^{2}$, as shown in Table \ref{table:table1}. 
Therefore, the condition $J_{\rm c}<J<J^{*}$ is satisfied, and the auto-oscillation was observed, which is shown in Fig. \ref{fig:fig2}(a).


Now let us consider the role of the feedback effect. 
Recall that the feedback effect modulates the demagnetization field as $4\pi M \to 4\pi M \{1+\nu [m_{x}(t-\tau)]^{2}\}$. 
As mentioned in Sec. \ref{sec:System description}, we focused on the positive $\nu$ case. 
This is because we intend to excite chaos by the feedback effect. 
For this purpose, the magnetization should show a sustainable dynamics, such as an auto-oscillation, while dynamics saturating to a fixed point, such as a magnetization switching, should be avoided. 
In our case, when $\nu$ is positive, the modulation of the demagnetization field by the feedback VCMA effect enhances the demagnetization field. 
Therefore, the threshold current density increases compared with that without the feedback effect, which makes the magnetization switching more difficult to happen. 
On the other hand, if $\nu$ is negative, the threshold current density becomes small, and the magnetization switching will be easier to appear. 
Therefore, expectation was that to use positive $\nu$ is suitable for exciting chaos in the present system. 
However, the numerical simulations show that, even though we choose the positive $\nu$, the magnetization switching occasionally occurs, as shown in Fig. \ref{fig:fig2}(c). 
This is the reason why we consider that the appearance of the transient chaos (magnetization switching) is non-trivial. 


\subsection{Switching caused by temporal change of the magnetic potential energy}
\label{sec:Switching caused by temporal change of the magnetic potential energy}



\begin{figure}
\centerline{\includegraphics[width=1.0\columnwidth]{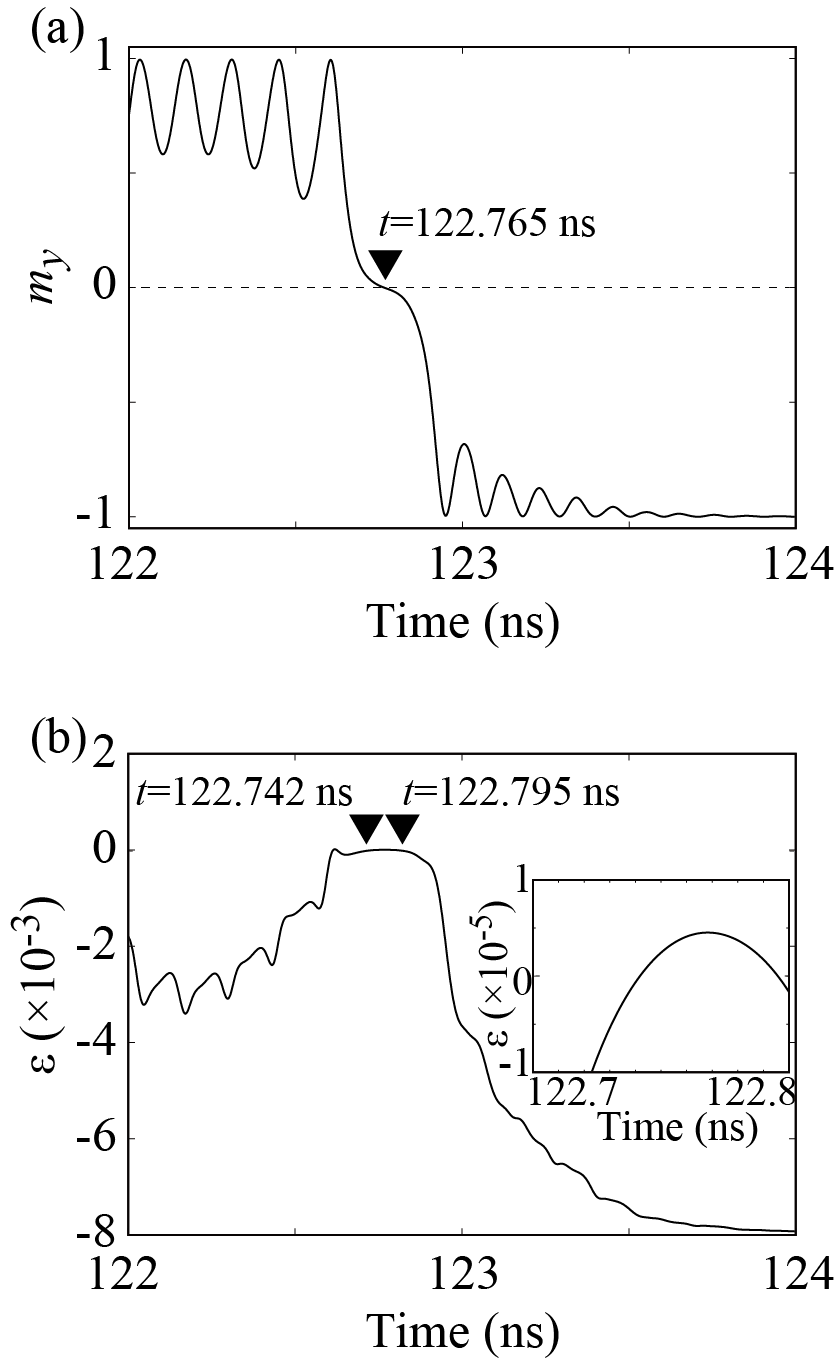}}
\caption{
            (a) Time evolution of $m_{y}$ near the switching. 
            (b) Time evolution of $\varepsilon$ near the switching. 
                 The inset shows an enlarged view near the region where $\varepsilon$ varies between positive and negative values.
         \vspace{-3ex}}
\label{fig:fig6}
\end{figure}



\begin{figure*}
\centerline{\includegraphics[width=2.0\columnwidth]{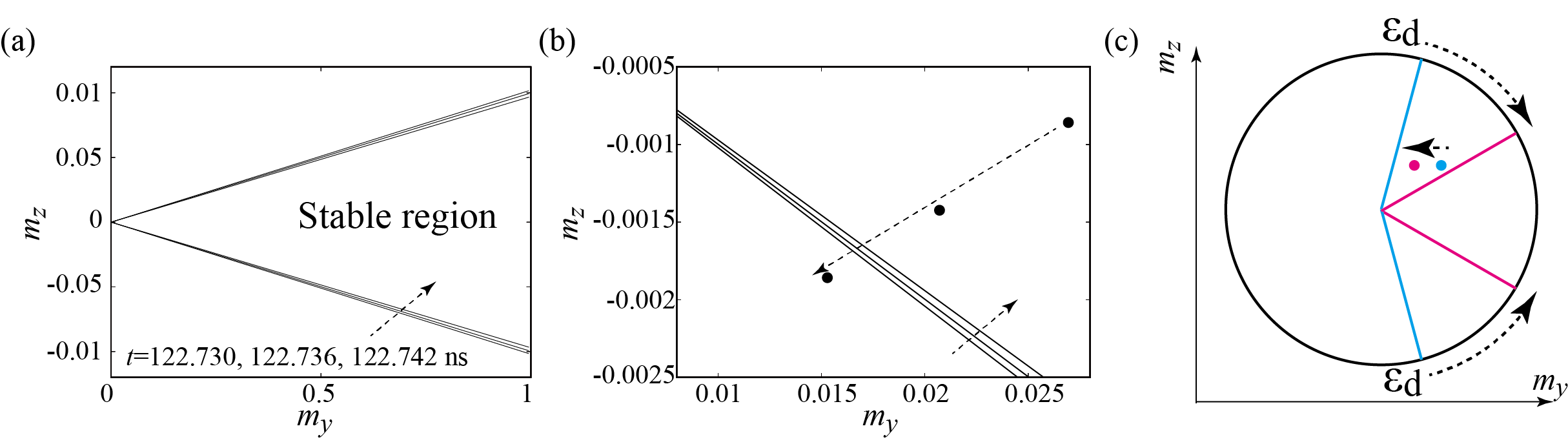}}
\caption{
            (a) Temporal change of constant energy lines of $\varepsilon=\varepsilon_{\rm d}$. 
                 A dotted arrow indicates the direction of the temporal change. 
                 As samples, $\varepsilon$ at $t=122.73$, $122.74$, and $122.75$ ns are chosen. 
            (b) Temporal change of the magnetization (dots). 
                 The sampling times are identical to those in Fig. \ref{fig:fig7}(a). 
                 The constant energy lines of $\varepsilon=\varepsilon_{\rm d}$ in Fig. \ref{fig:fig7}(a) are again shown to clarify that the magnetization crosses the lines because the directions of the temporal changes of the constant energy lines and the magnetization are opposite. 
            (c) Schematic illustration roughly describing the temporal changes of the constant energy line of $\varepsilon$ and the magnetization.
                The blue line and dot represent the constant energy line and the magnetization at an initial moment. 
                In the next stage, the constant energy line moves so that the stable region becomes narrow, while the magnetization moves to the direction to the saddle point, which are represented by purple line and dot. 
                When these temporal changes of the constant energy line and the magnetization occurs simultaneously in coincidence, the magnetization crosses the saddle point. 
         \vspace{-3ex}}
\label{fig:fig7}
\end{figure*}



To clarify the origin of the magnetization switching, it is helpful to compare this switching to the conventional auto-oscillation. 
As mentioned in Sec. \ref{sec:Why is the switching non-trivial?}, the trajectory of the conventional auto-oscillation stays inside the stable region, and thus, the energy density $E$ satisfies $E<E_{\rm d}$. 
Accordingly, the appearance of the magnetization switching in the presence of the feedback effect indicates that the magnetization crosses a constant energy line including the saddle point, even though the current density $J$ is smaller than the threshold current density. 
To confirm it, we introduce a normalized (dimensionless) energy density $\varepsilon$ defined as 
\begin{equation}
  \varepsilon
  =
  -\frac{k}{2}
  m_{y}^{2}
  +
  \frac{1+\nu [m_{x}(t-\tau)]^{2}}{2}
  m_{z}^{2}, 
  \label{eq:energy_delay}
\end{equation}
where $k=H_{\rm K}/(4\pi M)$. 
We note that Eq. (\ref{eq:energy_delay}) relates to Eq. (\ref{eq:energy}) via $E=4\pi M^{2}\varepsilon$ with $\nu \to 0$. 
In Eq. (\ref{eq:energy_delay}), the feedback term $m_{x}(t-\tau)$ is treated as a time-dependent parameter. 
Then, the saddle points of $\varepsilon$ still locate at $\mathbf{m}=\pm\mathbf{e}_{x}$, and the corresponding saddle point energy is $\varepsilon_{\rm d}=0$. 
We notice that the magnetization switching in the transient chaos accompanies the change of $\varepsilon$ from $\varepsilon<\varepsilon_{\rm d}$ to $\varepsilon>\varepsilon_{\rm d}$.  
For example, Fig. \ref{fig:fig6}(a) shows the time evolution of $m_{y}$ near the switching in Fig. \ref{fig:fig2}(c), where $m_{y}$ changes from positive to negative at $t=122.765$ ns. 
The time evolution of $\varepsilon$ near the switching is shown in Fig. \ref{fig:fig6}(b), where we confirm that $\varepsilon$ becomes positive in the range of $122.742 \le t \le 122.795$ ns, i.e., $\varepsilon$ around the magnetization switching indeed becomes larger than $\varepsilon_{\rm d}(=0)$ [see also the inset of Fig. \ref{fig:fig6}(b)]. 
These results indicate that $\varepsilon$ is no longer kept to be constant and can be larger than the saddle-point energy density when the transient chaos occurs. 
The result is in contrast to the conventional auto-oscillation, where $E$ is approximately constant and is in the region of $E<E_{\rm d}$. 
We also note that the temporal change of the energy density $\varepsilon$ is different from the conventional spin-transfer torque switching without feedback effect, where an averaged amplitude, as well as energy density, of the magnetization oscillation increases monotonically due to the energy injection by the work done by the spin-transfer torque.


Now let us investigate how the magnetization crosses the constant energy line including the saddle point. 
Note that a constant energy line of $\varepsilon$ changes temporally due to the presence of the feedback term. 
In other words, the stable and unstable regions are narrowed or expanded temporally, according to the oscillation of the feedback term $[m_{x}(t-\tau)]^{2}$.
We then notice that the switching occurs when the feedback term narrows the stable region.  
This can be confirmed from Fig. \ref{fig:fig7}(a), where we show the constant energy lines including the saddle points at various times ($t=122.730$, $122.736$, and $122.742$ ns) just before the magnetization switching. 
At the same time, we also notice that the magnetization locates close to the saddle point and its precession direction points from the stable state ($\mathbf{m}\parallel \mathbf{e}_{y}$) to the saddle point ($\mathbf{m}\parallel \mathbf{e}_{x}$). 
This can be confirmed from Fig. \ref{fig:fig7}(b), where we show the magnetization directions by dots and again show the constant energy lines including the saddle point by solid lines. 
The dotted arrows indicate the directions of their temporal changes. 
It is shown that the constant energy line and the magnetization move in the opposite directions. 
One may consider that the change of the constant energy line is small compared to that of the magnetization. 
Recall, however, that the stable regions are narrow due to the large demagnetization field and the constant energy lines inside the regions are densely distributed; thus, a small change of the constant energy line including the saddle point significantly affects the alignment of the other constant energy lines. 
As a result, the magnetization crosses the constant energy line including the saddle point. 
Hence, the magnetization enters the unstable region, precesses around the demagnetization field, and relaxes to the switched state. 


Summarizing these results, we consider that the magnetization switching occurs when the following conditions are coincidentally satisfied. 
First, the magnetization should be located near the saddle point, or equivalently, far away from the easy axis. 
It occurs approximately periodically because the magnetization precesses around the easy axis. 
Second, the precession direction points to the saddle point, which also occurs almost periodically. 
Third, the modulation of the demagnetization field narrows the stable region. 
Recall that the stable region repeats narrowing and expanding because the feedback term $[m_{x}(t-\tau)]^{2}$ oscillates. 
In Fig. \ref{fig:fig7}(c), we show a schematic illustration, which roughly summarizes a situation satisfying these conditions. 
We emphasize that these conditions should be satisfied simultaneously. 
For example, even if the modulation of the demagnetization field narrows the stable region and as a result, the magnetization is located at a point outside a constant energy line including the saddle point, the switching will not occur if the location is close to the stable state.  
This is because a relatively long time is necessary for switching in this case, however, before that, the stable region may expand according to the temporal change of the feedback term. 
Therefore, the magnetization switching rarely occurs and/or strongly depends on the values of the parameters and the initial condition. 
This leads to that, although chaos is excited over a wide range of the parameter $\nu$, transient chaos is also observed in some cases. 


\section{Conclusion}
\label{sec:Conclusion}

In summary, theoretical investigation is carried out of the magnetization dynamics in a three-terminal STO modulated by the feedback VCMA effect. 
Both the temporal analysis of the magnetization dynamics and the statistical evaluation of the Lyapunov exponent indicated the existence of chaos in the STO. 
At the same time, however, the transient chaos was also observed, where the magnetization initially shows chaotic behavior but finally switches its direction and saturates to a fixed point. 
This transient dynamics was unexpected because we chose the sign of the feedback VCMA effect so that the switching current of the magnetization increases, hence avoiding the switching and sustaining chaos. 
By analyzing the temporal change of the magnetic anisotropy energy, it was suggested that the transient chaos (magnetization switching) occurs when the feedback effect narrows the stable region of the potential while the magnetization precesses to the direction of the saddle point simultaneously. 
These makes it possible for the magnetization to cross a constant energy line including the saddle point and switches the magnetization direction. 
As a result, while chaos was observed over a wide range of the parameters, the transient chaos also appeared in some cases. 


\section*{Acknowledgements}

The work is supported by JSPS KAKENHI Grants No. 20H05655 and No. 24K01336. 
The author is grateful to Takayuki Nozaki and Takehiko Yorozu for valuable discussion. 


\appendix


\section{Averaged LLG equation of in-plane magnetized ferromagnet}
\label{sec:AppendixA}


Here, we describe the derivation of the threshold current density $J^{*}$ in Sec. \ref{sec:Analysis of transient chaos} from the averaged LLG equation \cite{bertotti09}. 
The basic idea of the averaged LLG equation is as follows \cite{bertotti04,bertotti05,bertotti06,bertotti09,newhall13}. 
As implied in Fig. \ref{fig:fig5}(a), we can cover a unit sphere by constant energy lines. 
Therefore, the magnetization direction on this sphere can be identified by specifying the value of the energy and a phase along a constant energy line. 
Note that a constant energy line is identical to the solution of the LL equation, $d\mathbf{m}/dt=-\gamma\mathbf{m}\times\mathbf{H}$. 
For example, in the absence of the spin-transfer and damping torques, the solution of the LL equation ($d\mathbf{m}/dt=-\gamma \mathbf{m}\times\mathbf{H}$) of the in-plane magnetized ferromagnet is ($\varphi_{0}$ is the phase giving the initial condition) \cite{taniguchi17}, 
\begin{equation}
  m_{x}
  =
  \sqrt{
    1
    +
    \frac{2E}{MH_{\rm K}}
  }
  {\rm sn}
  \left[
    \frac{4 \mathsf{K}(k)}{\tau(E)}
    t
    +
    \varphi_{0},
    k
  \right],
  \label{eq:mx}
\end{equation}
\begin{equation}
  m_{y}
  =
  \sqrt{
    \frac{4\pi M-2E/M}{H_{\rm K}+4\pi M}
  }
  {\rm dn}
  \left[
    \frac{4 \mathsf{K}(k)}{\tau(E)}
    t
    +
    \varphi_{0},
    k
  \right],
  \label{eq:my}
\end{equation}
\begin{equation}
  m_{z}
  =
  \sqrt{
    \frac{H_{\rm K}+2E/M}{H_{\rm K}+4\pi M}
  }
  {\rm cn}
  \left[
    \frac{4\mathsf{K}(k)}{\tau(E)}
    t
    +
    \varphi_{0},
    k
  \right],
  \label{eq:mz}
\end{equation}
where ${\rm sn}(u,k)$, ${\rm dn}(u,k)$, and ${\rm cn}(u,k)$ are the Jacobi elliptic functions with the modulus $k$, whereas $\mathsf{K}(k)$ is the first kind of complete elliptic integral. 
The modulus is 
\begin{equation}
  k
  =
  \sqrt{
    \frac{4\pi M (H_{\rm K}+2E/M)}{H_{\rm K}(\pi M-2E/M)}
  }.
  \label{eq:modulus}
\end{equation}
The oscillation period $\tau(E)$ is related to the frequency $f(E)$ of the oscillation along a constant energy line of $E$ as $\tau(E)=1/f(E)$, where,  
\begin{equation}
  f(E)
  =
  \frac{\gamma \sqrt{H_{\rm K}(4\pi M-2E/M)}}{4\mathsf{K}(k)}. 
  \label{eq:frequency}
\end{equation}


In the auto-oscillation state, the spin-transfer torque balances with the damping torque, and $E$ is kept to be approximately constant. 
In fact, if $E$ changes temporarily, the magnetization will relax or switch its direction, and the auto-oscillation will not appear. 
Therefore, the magnetization dynamics in the auto-oscillation state is well approximated by Eqs. (\ref{eq:mx}), (\ref{eq:my}), and (\ref{eq:mz}). 
In other words, $E$ changes slowly while the phase changes rapidly. 
Note that the change of $E$ is described by the following equation, which can be derived from the LLG equation, Eq. (\ref{eq:LLG}), by using the relation $dE/dt=-M \mathbf{H}\cdot (d \mathbf{m}/dt)$; 
\begin{equation}
\begin{split}
  \frac{dE}{dt}
  =&
  \gamma M 
  \frac{\hbar\vartheta J}{2eMd}
  \left[
    \mathbf{e}_{y}
    \cdot
    \mathbf{H}
    -
    \left(
      \mathbf{m}
      \cdot
      \mathbf{e}_{y}
    \right)
    \left(
      \mathbf{m}
      \cdot
      \mathbf{H}
    \right)
  \right]
\\
  &-
  \alpha
  \gamma
  M
  \left[
    \mathbf{H}^{2}
    -
    \left(
      \mathbf{m}
      \cdot
      \mathbf{H}
    \right)^{2}
  \right],
  \label{eq:dEdt}
\end{split}
\end{equation}
where we neglect the terms on the order of $\alpha^{2}(\ll 1)$. 
In fact, the averaging technique of the LLG equation is valid when $\alpha \ll 1$ because of the following reason. 
As mentioned, the spin-transfer and damping torques averaged over a period are balanced each other in an auto-oscillation state. 
Therefore, an averaged energy density $E$ is also kept to be constant, and the oscillation trajectory is approximated by a constant energy line of $E$. 
Note, however, that an instantaneous strength of the spin-transfer and damping torques might be different, even if their averages are balanced, depending on various factors such as the angle dependence of the spin-transfer torque and the magnetic anisotropy. 
Thus, an instantaneous position of the magnetization might also differ from the constant energy line of $E$ if the value of $\alpha$ is large. 
Regarding these points, the averaged LLG equation is useful to analyze an auto-oscillation state when the value of $\alpha$ is small. 
Since the magnitude of the damping torque is roughly given by $\alpha |\mathbf{H}|$, the strength of the spin-transfer torque in the unit of magnetic field, $\hbar\vartheta J/(2eMd)$, should also be on the order of $\alpha |\mathbf{H}|$ for realizing the balance between these torques. 
Therefore, the terms such as $\alpha \hbar\vartheta J/(2eMd)$ should also be neglected as higher order terms of $\alpha$. 

According to the above discussion, the average of the temporal change of $E$ over the oscillation period is zero when the magnetization is in an auto-oscillation state; 
\begin{equation}
  \oint
  dt 
  \frac{dE}{dt}
  =
  0, 
  \label{eq:dEdt_ave1}
\end{equation}
where the integral is over a period of a constant energy line, $[0,\tau(E)]$. 
Substituting Eqs. (\ref{eq:mx}), (\ref{eq:my}), and (\ref{eq:mz}) into the right-hand-side of Eq. (\ref{eq:dEdt}) and performing an integral, Eq. (\ref{eq:dEdt_ave1}) becomes 
\begin{equation}
\begin{split}
&
  \frac{\pi\hbar\vartheta J(H_{\rm K}+2E/M)}{ed \sqrt{H_{\rm K}(H_{\rm K}+4\pi M)}}
\\
  &\ \ \ \ 
  -
  4 \alpha M 
  \sqrt{
    \frac{4\pi M-2E/M}{H_{\rm K}}
  }
  \left[
    \frac{2E}{M}
    \mathsf{K}(k)
    +
    H_{\rm K}
    \mathsf{E}(k)
  \right]
\\
  &=0,
  \label{eq:dEdt_ave2}
\end{split}
\end{equation}
where $\mathsf{E}(k)$ is the second kind of complete elliptic integral. 
Equation (\ref{eq:dEdt_ave2}) can be regarded as an equation determining the current density $J$ to make the balance between spin-transfer and damping torques and keep an auto-oscillation along a constant energy line of $E$. 
In other words, by specifying the value of $E$, we can estimate the current density necessary to excite an auto-oscillation, where the trajectory is well approximated by Eqs. (\ref{eq:mx}), (\ref{eq:my}), and (\ref{eq:mz}). 
In this sense, the current density $J$ satisfying Eq. (\ref{eq:dEdt_ave2}) can be regarded as a function of $E$. 
The critical and threshold current densities are then given by 
\begin{equation}
  J_{\rm c}
  =
  \lim_{E \to E_{\rm min}}
  J(E),
\end{equation}
\begin{equation}
  J^{*}
  =
  \lim_{E \to  E_{\rm d}}
  J(E). 
\end{equation}
According to these definitions, $J_{\rm c}$ and $J^{*}$ are the current densities necessary to make balance between the spin-transfer and damping torques along the constant energy lines including the stable and saddle points, respectively. 
For typical in-plane magnetized ferromagnets, $H_{\rm K}\ll 4\pi M$ is often satisfied. 
In this case, $J^{*}/J_{\rm c}\simeq 4/\pi \simeq 1.27$. 
However, using various factors, including the VCMA effect, it may be possible to reduce the magnitude of the demagnetization field ($4\pi M$) significantly and make $J^{*}/J_{\rm c}<1$. 


The critical and threshold current densities can be used to classify the magnetization dynamics. 
When $J/J_{\rm c}<1$, the magnetization stays near the easy axis because the spin-transfer torque does not overcome the damping torque near the stable state. 
When $J/J_{\rm c}>1$, the magnetization can move from the easy axis. 
Then, if $J^{*}/J_{\rm c}>1$ is also satisfied, the magnetization shows an auto-oscillation when $J/J^{*}<1$ is also satisfied. 
The oscillation trajectory is well described by Eqs. (\ref{eq:mx}), (\ref{eq:my}), and (\ref{eq:mz}), as mentioned above. 
When $J/J^{*}>1$, the magnetization switches its direction. 
If $J^{*}/J_{\rm c}<1$, on the other hand, the magnetization switches its direction without showing an auto-oscillation. 
In the present study, since we are interested in exciting chaos, we choose the values of the parameters to satisfy the conditions, $J/J_{\rm c}>1$, $J/J^{*}<1$, and $J^{*}/J_{\rm c}>1$, at least in the absence of the feedback effect. 



%


\end{document}